# THEORY OF MAGNETIC CIRCULAR DICHROISM SPECTROSCOPY TECHNIQUE IN REFLECTED LIGHT USING AN ADDITIONAL MIRROR

**Yu. V. Markin[a,*] and Z. E. Kun'kova[a,**]**

[a]*Kotelnikov Institute of Radio Engineering and Electronics, Russian Academy of Sciences (Fryazino Branch),*

*Fryazino, Moscow oblast, 141190 Russia*

*\*e-mail: markin@fireras.su*

*\*\*e-mail: zek@fireras.su*

The theory of the method of magnetic circular dichroism spectroscopy in reflected light (RMCD) has been developed using a simplified scheme without an analyzer, with the use of an additional mirror in the optical system and with the application of the method of phase modulation of the light wave using a photoelastic modulator. An additional mirror reflector, together with the mirror surface of the sample, forms a double mirror located between the pole pieces of a powerful laboratory electromagnet. The presence of an additional mirror leads to mixing of the amplitude ($\Delta R/R$) and phase ($\Delta \phi$) RMCD-components in the amplitudes of the signals at the first and second harmonics of the phase modulation. It is shown that the RMCD spectroscopy technique using a simplified scheme and an additional mirror requires the synchronous measurement of three electrical signals proportional to the amplitudes of the corresponding terms in the expression for the total intensity of monochromatic radiation detected by the photodetector: a constant signal $V_{dc}$ and high-frequency amplitudes $V_f$ and $V_{2f}$ at phase modulation frequencies $f$ and $2f$. The ratios $V_f/V_{dc}$ and $V_{2f}/V_{dc}$ determine a system of linear equations for $\Delta R/R$ and $\Delta \phi$ with known ellipsometric parameters $\tan\Psi$ and $\Delta$ of the additional mirror, which are found either from additional measurements using an ellipsometer or directly (in sity) using the RMCD spectrometer itself.

Reflectance magnetic circular dichroism (RMCD) spectroscopy is based on the technique of measuring the polar magneto-optical Kerr effect at near-normal incidence of polarized monochromatic radiation on the sample surface [1,2]. In polar geometry, the external magnetic field vector is perpendicular to the plane of the sample [3]. In this case, to implement polar geometry, the sample is placed in the gap between the pole pieces of an electromagnet and oriented so that its surface is parallel to the surface of the pole tips. One of the pole pieces has a narrow (usually a few millimeters in diameter) axial hole through which light rays incident on and reflected from the sample surface pass [1,2]. When using powerful laboratory electromagnets, such a hole must be significantly long (on the order of tens of centimeters), making RMCD spectroscopy difficult.

In [4], an optical scheme for magneto-optical spectroscopy is proposed, in which an additional mirror reflector is introduced, which, together with the mirror surface of the sample, forms a double mirror. This scheme eliminates the need for a pole piece with an axial hole [1,2].

The theory [4] is built for a traditional magneto-optical spectroscopy scheme containing an analyzer in the spectrometer's optical path. In works [5, 6], a simpler method for measuring RMCD spectra was used, in which the analyzer was absent. This simplifies the organization of experiments and the setup of the measurement system, and also improves the sensitivity and accuracy of measurements. This paper presents a simplified RMCD spectroscopy technique using a double mirror located between the pole pieces of a powerful laboratory electromagnet.

Let's show that RMCD spectroscopy is possible with measurements using the simplified *P–PEM–M–Sp–M* scheme (Fig. 1a). In the optical scheme, *P* is a linear polarizer; *PEM* is a photoelastic modulator; *M* is an additional mirror located parallel to the *y*-axis, on which the primary light beam, propagating along the *z*-axis, is incident at an angle close to 45°; *Sp* is a sample, the illuminated surface of which is perpendicular to the *x*-axis and the magnetic induction vector $\mathbf{B}_0$ (Fig. 1b).

When polarized light passes through optical elements located on the optical path of its propagation after the polarizer, as well as when the beam is reflected from the surface of the additional mirror and the surface of the sample, a change in the polarization state of the light radiation occurs. To analyze the time dependence of the intensity of the light wave recorded by the photodetector, we will trace this change along the path of light propagation from the polarizer to the optical radiation receiver, using the Johnson matrix formalism [7]. For this purpose, we introduce a right-hand Cartesian coordinate system *xyz*, coinciding with the modulator's own axes (Fig. 1b). The *x*-axis is aligned with the optical axis *OO′* of the PEM, the *y*-axis is perpendicular to the modulator's main plane passing through the light beam, and the *z*-axis is parallel to the light beam in the direction of its propagation. The active element of the modulator's optical head is oriented such that its face, which receives the beam, is perpendicular to the PEM's main plane. In this case, the light beam will fall normally on the modulator's surface.

Typically, the polarizer's transmission axis is oriented relative to the x-axis at an angle of 45°. Let unpolarized monochromatic light with intensity $I_0(\lambda)$ (λ – wavelength of light radiation) fall on the polarizer. Then, the Jones column vector of the light field after the polarizer before the PEM has the form:

$$\mathbf{E}_0 = \sqrt{\frac{I_0(\lambda)}{2}} \begin{pmatrix} 1 \\ 1 \end{pmatrix}. \tag{1}$$

In the modulator's own axes, its Jones matrix is equal to

$$\mathbf{T} = \begin{pmatrix} e^{i\delta(t)} & 0 \\ 0 & 1 \end{pmatrix}, \tag{2}$$

where *i* is the imaginary unit, and

$$\delta(t) = \delta_0 \sin(\omega_{PEM} t). \tag{3}$$

Here $\delta_0$, and $\omega_{PEM}$ are the amplitude and frequency of the phase modulation PEM, respectively, and *t* is the current time. The Jones matrices describing the change in the polarization state of the rays during the first and second reflections from the additional mirror *M* are as follows:

$$\mathbf{M}_1 = \begin{pmatrix} r_p(\alpha/2) & 0 \\ 0 & r_s(\alpha/2) \end{pmatrix} = \begin{pmatrix} |r_p(\alpha/2)| e^{i\varphi_p(\alpha/2)} & 0 \\ 0 & |r_s(\alpha/2)| e^{i\varphi_s(\alpha/2)} \end{pmatrix},$$

$$\mathbf{M}_2 = \begin{pmatrix} r_p(\beta/2) & 0 \\ 0 & r_s(\beta/2) \end{pmatrix} = \begin{pmatrix} |r_p(\beta/2)| e^{i\varphi_p(\beta/2)} & 0 \\ 0 & |r_s(\beta/2)| e^{i\varphi_s(\beta/2)} \end{pmatrix}, \tag{4}$$

where $r_p$ and $r_s$ are the complex Fresnel coefficients for the corresponding *p*- and *s*-components of the electric field strength of the reflected light wave, $\varphi_p$ and $\varphi_s$ are the phase changes of the corresponding *p*- and *s*-components upon reflection from the mirror surface. In the polar geometry under consideration (Fig. 1b), the angle $\gamma$ is small, and in further calculations we will make the approximation $\alpha = \beta$. Then

$$\mathbf{M}_1 = \mathbf{M}_2 = \mathbf{M} = \begin{pmatrix} |r_p| e^{i\varphi_p} & 0 \\ 0 & |r_s| e^{i\varphi_s} \end{pmatrix}. \tag{5}$$

Let's consider the change in the polarization state of a light beam upon reflection from a sample in a circular basis, in which the light wave is represented as a superposition of two orthogonal circularly polarized components. Let the complex Fresnel coefficients $r_R = |r_R| e^{i\phi_R}$ and $r_L = |r_L| e^{i\phi_L}$ describe the reflection of waves with right- and left-circular polarizations, respectively. Then the Jones matrix for normal reflection from the sample surface has the form:

$$\mathbf{S} = \begin{pmatrix} |r_R| e^{i\phi_R} & 0 \\ 0 & |r_L| e^{i\phi_L} \end{pmatrix} \tag{6}$$

The Jones vector of the light field at the input window of the photodetector *PHD* (Fig. 1a) is determined by the matrix product

$$\mathbf{E}_D = \mathbf{M} \cdot \mathbf{Q}_{CD} \cdot \mathbf{S} \cdot \mathbf{Q}_{DC} \cdot \mathbf{M} \cdot \mathbf{T} \cdot \mathbf{E}_0 \tag{7}$$

where $\mathbf{Q}_{DC}$ and $\mathbf{Q}_{CD}$ are the matrix of the transition from a linear basis to a right circular basis, and the matrix of the inverse transition from a circular basis to a linear basis, respectively:

$$\mathbf{Q}_{DC} = \frac{1}{\sqrt{2}} \begin{pmatrix} 1 & -i \\ -i & 1 \end{pmatrix}, \quad \mathbf{Q}_{CD} = \frac{1}{\sqrt{2}} \begin{pmatrix} 1 & i \\ i & 1 \end{pmatrix}. \tag{8}$$

Then the intensity of light that has passed through the optical system under consideration, measured by the photodetector, is determined by the expression

$$\begin{aligned} I(t,\lambda) \propto \mathbf{E}_D \cdot \mathbf{E}_D^* = \frac{I_0(\lambda)}{4} |r_s|^4 R(\lambda) \Big\{ &2(1 + \tan^4 \Psi) + \\ &J_0(\delta_0) \tan \Psi \Big[ (1 + \tan^2 \Psi) \sin \Delta \cdot \frac{\Delta R}{R} + 2(1 - \tan^2 \Psi) \cos \Delta \cdot \Delta \phi \Big] + \\ &2 J_1(\delta_0) \tan \Psi \Big[ (1 + \tan^2 \Psi) \cos \Delta \cdot \frac{\Delta R}{R} - 2(1 - \tan^2 \Psi) \sin \Delta \cdot \Delta \phi \Big] \sin(\omega_{PEM} t) + \\ &2 J_2(\delta_0) \tan \Psi \Big[ (1 + \tan^2 \Psi) \sin \Delta \cdot \frac{\Delta R}{R} + 2(1 - \tan^2 \Psi) \cos \Delta \cdot \Delta \phi \Big] \cos(2\omega_{PEM} t) \Big\}, \end{aligned} \tag{9}$$

where $\tan \Psi$ and $\Delta$ are the ellipsometric parameters of the additional mirror: $\tan \Psi = |r_p|/|r_s|$, $\Delta = \varphi_p - \varphi_s$; $\Delta R = R_L - R_R$, $R = (R_L + R_R)/2$, $R_L = |r_L|^2$, $R_R = |r_R|^2$, $\Delta \phi = \phi_L - \phi_R$, $R_L$ ($\phi_L$) and $R_R$ ($\phi_R$) are the reflection coefficients (phase shifts) for light waves with left and right circular polarizations, respectively; $J_0$, $J_1$, $J_2$ are the Bessel functions of zero, first and second order, respectively. Expression (9) is obtained using the Fourier series expansion of the functions $\sin[\delta_0 \sin(\omega_{PEM} t)]$ and $\cos[\delta_0 \sin(\omega_{PEM} t)]$ taking into account the conditions $\Delta R \ll 1$ and $\Delta \phi \ll 1$. Note that if the amplitude anisotropy of the reflector $M$ is absent ($\tan \Psi = 1$), and the difference in phase shifts $\Delta = 180°$, then expression (9) goes over to the well-known one [2,5].

From (9) it follows that the RMCD spectroscopy technique using an additional mirror requires the synchronous measurement of three electrical signals proportional to the amplitudes of the corresponding terms in the expression for the total intensity $I(t,\lambda)$:

$$\begin{aligned} V_{dc}(\lambda) \propto & \\ &\frac{I_0(\lambda)}{4} |r_s|^4 R(\lambda) \Big\{ 2(1 + \tan^4 \Psi) + \\ &J_0(\delta_0) \tan \Psi \Big[ (1 + \tan^2 \Psi) \sin \Delta \cdot \frac{\Delta R}{R} + 2(1 - \tan^2 \Psi) \cos \Delta \cdot \Delta \phi \Big] \Big\}, \end{aligned} \tag{10}$$

$$V_f(\lambda) \propto$$
$$\frac{I_0(\lambda)}{2} |r_s|^4 R(\lambda) J_1(\delta_0) \tan \Psi \left[ (1+\tan^2 \Psi)\cos\Delta \cdot \frac{\Delta R}{R} - 2(1-\tan^2 \Psi)\sin\Delta \cdot \Delta\phi \right] \quad (11)$$

and

$$V_{2f}(\lambda) \propto$$
$$\frac{I_0(\lambda)}{2} |r_s|^4 R(\lambda) J_2(\delta_0) \tan \Psi \left[ (1+\tan^2 \Psi)\sin\Delta \cdot \frac{\Delta R}{R} + 2(1-\tan^2 \Psi)\cos\Delta \cdot \Delta\phi \right]. \quad (12)$$

From expressions (10) – (12) it follows that with a phase delay amplitude of $\delta_0 = 0.383\lambda$ (i.e., with $J_0(\delta_0) = 0$) the ratio of the high-frequency amplitudes at frequencies $f$ and $2f$ ($f = 2\pi\omega_{PEM}$) to the constant signal

$$\begin{aligned}
\frac{V_f}{V_{dc}} &= \frac{\tan\Psi}{1+\tan^4\Psi} J_1(\delta_0) \left[ (1+\tan^2\Psi)\cos\Delta \cdot \frac{\Delta R}{R} - 2(1-\tan^2\Psi)\sin\Delta \cdot \Delta\phi \right] \\
\frac{V_{2f}}{V_{dc}} &= \frac{\tan\Psi}{1+\tan^4\Psi} J_2(\delta_0) \left[ (1+\tan^2\Psi)\sin\Delta \cdot \frac{\Delta R}{R} + 2(1-\tan^2\Psi)\cos\Delta \cdot \Delta\phi \right]
\end{aligned} \quad (13)$$

determine a system of linear equations for $\Delta R/R$ and $\Delta\phi$ with known parameters $\tan\Psi$ and $\Delta$, which are found from additional measurements using an ellipsometer [4].

We will show that, when measuring using the *P–PEM–M–SP–M–A* scheme in the absence of a magnetic field, the spectral dependences can be found directly (in situ) using the spectrometer itself. Here, *A* is a linear analyzer. The Jones vector of the light field at the input window of the photodetector is determined by the matrix product

$$\mathbf{E}_D = \mathbf{A} \cdot \mathbf{R} \cdot \mathbf{M} \cdot \mathbf{S}_0 \cdot \mathbf{M} \cdot \mathbf{T} \cdot \mathbf{E}_0, \quad (14)$$

where $\mathbf{S}_0$, $\mathbf{R}$ and $\mathbf{A}$ are the Jones matrices describing the normal reflection of light from the sample surface in the absence of a magnetic field, the transition from the original coordinate system to the analyzer's own axes and the analyzer itself, respectively:

$$\mathbf{S}_0 = \sqrt{R_S} \begin{pmatrix} -1 & 0 \\ 0 & 1 \end{pmatrix}, \quad \mathbf{R} = \begin{pmatrix} \cos\varphi & \sin\varphi \\ -\sin\varphi & \cos\varphi \end{pmatrix}, \quad \mathbf{A} = \begin{pmatrix} 1 & 0 \\ 0 & 0 \end{pmatrix}. \quad (15)$$

Here $R_S$ is the reflectivity of light from the sample surface, and $\varphi$ is the analyzer azimuth. Let's find the intensity of light incident on the photodetector for three values of the analyzer azimuth $\varphi$:

$$I_P(\lambda) = \frac{I_0(\lambda)}{2}|r_s|^4 R_S \tan^4 \Psi, \quad \varphi = 0°, \tag{16}$$

$$I_S(\lambda) = \frac{I_0(\lambda)}{2}|r_s|^4 R_S, \quad \varphi = 90°, \tag{17}$$

$$I_{SP}(t,\lambda) = \frac{I_0(\lambda)}{4}|r_s|^4 R_S \{1 + \tan^4 \Psi + 2\tan^2 \Psi [J_0(\delta_0)\cos(2\Delta) - \\ 2J_1(\delta_0)\sin(2\Delta)\sin(\omega_{PEM}t) + 2J_2(\delta_0)\cos(2\Delta)\cos(2\omega_{PEM}t)]\}, \quad \varphi = -45°. \tag{18}$$

In the absence of phase modulation ($\delta_0 = 0$) expression (18) is simplified:

$$I_{SP}(\lambda) = \frac{I_0(\lambda)}{4}|r_s|^4 R_S \left[1 + \tan^4 \Psi + 2\tan^2 \Psi \cos(2\Delta)\right]. \tag{19}$$

From (16), (17) and (19) it follows that the method for determining in situ the $\tan\Psi$ and $\Delta$ of a double mirror requires performing three consecutive scans over wavelengths in a given range with constant values of the luminous flux of the primary light source and the light sensitivity of the photodetector. Then, the electrical signals of the light detector, proportional to the corresponding intensities (16), (17), (19), can be determined by one of two methods: the electronic direct current method or the electronic alternating current method, which involves periodic interruption of the light beam. In this case, measurements are performed using a single device: either a direct current voltmeter (static photometric ellipsometry) or a synchronous amplifier (dynamic photometric ellipsometry).

On the other hand, with a constant analyzer azimuth $\varphi = -45°$, according to relation (18), to determine in situ $\tan\Psi$ and $\Delta$ of a double mirror, synchronous measurements of three electrical signals proportional to the amplitudes of the corresponding terms in the expression for the total intensity are sufficient with a single wavelength scan: $V_{dc}(\lambda)$, $V_f(\lambda)$ and $V_{2f}(\lambda)$.

In conclusion, we note that with standard RMCD spectroscopy using the complete *P – PEM – SP – A* scheme and the phase modulation method, the signal amplitudes at the first and second harmonics are proportional to the amplitude $\Delta R/R$ and phase $\Delta\phi$ RMCD components [1,2]: $V_f \propto \Delta R/R$, $V_{2f} \propto \Delta\phi$. With RMCD spectroscopy using the incomplete *P–PEM–SP* scheme, the light intensity received by the photodetector contains exclusively odd harmonics [2], i.e., only the spectral dependence of the amplitude RMCD-component can be found here. If necessary, the spectral dependence of the phase RMCD-component can be reconstructed from the amplitude one using the dispersion relation [8].

Reflection of light from a mirror with a metal coating is characterized by amplitude-phase anisotropy, which depends on the type of metal, the type of coating and the angle of incidence [7]. In this case, with the exception of normal incidence, the polarization state of the beam reflected from the mirror reflector will change. From relations (11) and (12) it is evident that in RMCD spectroscopy using an incomplete scheme, the mirror causes mixing of signals corresponding to the RMCD-components: $V_f \propto a \cdot \Delta R/R + b \cdot \Delta\phi$, $V_{2f} \propto c \cdot \Delta R/R + d \cdot \Delta\phi$, where the proportionality coefficients $a$, $b$, $c$, and $d$ depend on the ellipsometric parameters of the mirror $\tan\Psi$ and $\Delta$. In other words, the simplified $P$–$PEM$–$M$–$Sp$–$M$ scheme implicitly becomes complete.

FUNDING

This study was performed within the framework of a State Task of Kotelnikov Institute of Radio Engineering and Electronics, Russian Academy of Sciences.

9...

**FIGURE CAPTIONS**

**Fig. 1**. Simplified block diagram of an RMCD spectrometer with an additional mirror (a) and a diagram of the arrangement of a double mirror between the pole pieces S, N of a powerful electromagnet, realizing the geometry of the polar Kerr effect with normal incidence of a light beam on the surface of a sample (b).

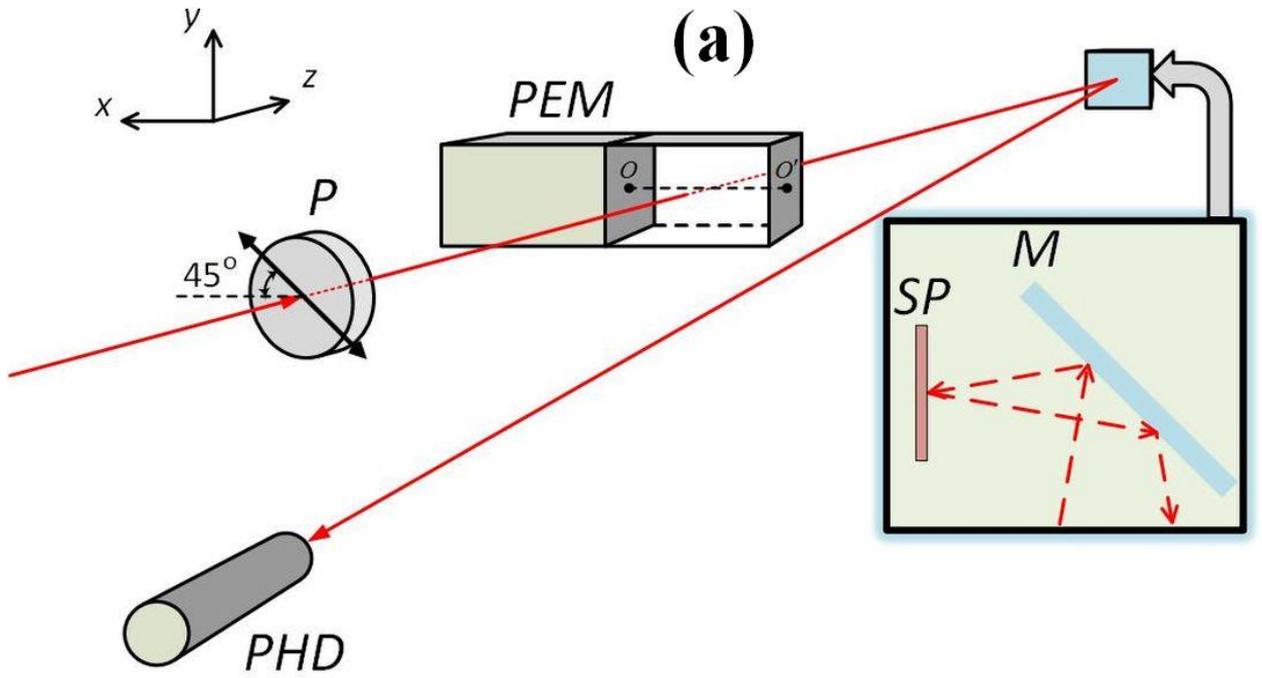

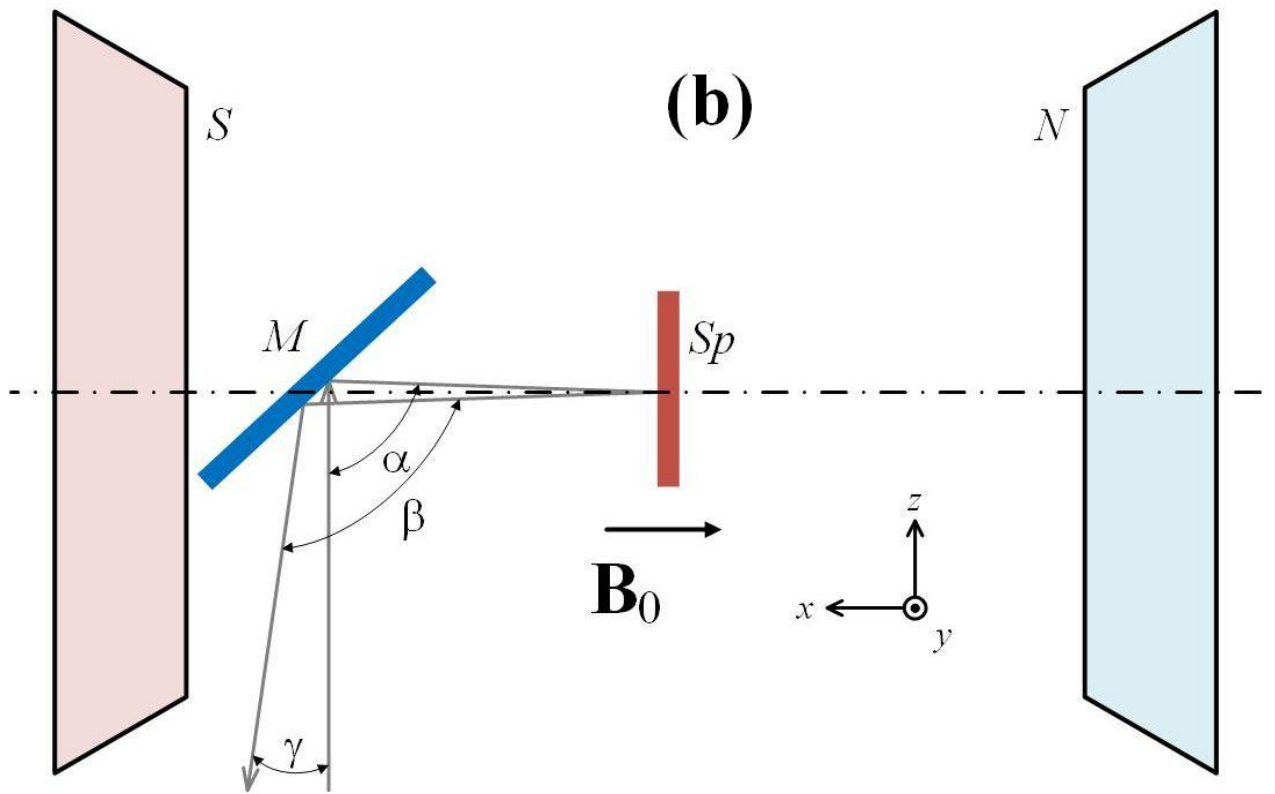